\documentclass[preprint,aps]{revtex4}

\usepackage{graphicx}% Include figure files
\usepackage{dcolumn}% Align table columns on decimal point
\usepackage{bm}% bold math

\begin{document}

\preprint{APS/123-QED}

\title{Energy scales of Lu$_{1-x}$Yb$_x$Rh$_2$Si$_2$ by means of thermopower investigations}

\author{U. K\"ohler}
\email{koehler@cpfs.mpg.de}
\author{N. Oeschler}
\author{F. Steglich}
\affiliation{Max Planck Institute for Chemical Physics of Solids,
D-01187 Dresden, Germany}%
\author{S. Maquilon}
\author{Z. Fisk}
\affiliation{Department of Physics and Astronomy, University of California, Irvine, CA 92697, USA}%

\date{\today}% It is always \today, today,
             %  but any date may be explicitly specified

\begin{abstract}
We present the thermopower $S(T)$ and the resistivity $\rho(T)$ of
Lu$_{1-x}$Yb$_x$Rh$_2$Si$_2$ in the temperature range 3~K $< T <
300$~K. $S(T)$ is found to change from two minima for dilute systems
($x < 0.5$) to a single large minimum in pure YbRh$_2$Si$_2$. A
similar behavior has also been found for the magnetic contribution
to the resistivity $\rho_\mathrm{mag}(T)$. The appearance of the
low-$T$ extrema in $S(T)$ and $\rho_\mathrm{mag}(T)$ is attributed
to the lowering of the Kondo scale $k_\mathrm{B}T_\mathrm{K}$ with
decreasing $x$. The evolution of the characteristic energy scales
for both the Kondo effect and the crystal electric field splitting
$\Delta _{\mathrm{CEF}}$ are deduced. An extrapolation of
$T_\mathrm{K}$ to $x=1$ allows to estimate the Kondo temperature of
YbRh$_2$Si$_2$ to 29~K. For pure YbRh$_2$Si$_2$, $T_\mathrm{K}$ and
$\Delta _{\mathrm{CEF}}/k_\mathrm{B}$ lie within one order of
magnitude and thus the corresponding extrema merge into one single
feature.
\end{abstract}

% PACS, the Physics and Astronomy
                             % Classification Scheme.
%\keywords{Suggested keywords}%Use showkeys class option if keyword
                              %display desired
\maketitle

\section{Introduction}

YbRh$_2$Si$_2$ is a stoichiometric heavy-fermion (HF) metal with an
extremely low antiferromagnetic ordering temperature of
70~mK.\cite{YRS-00-1} It crystallizes in the tetragonal
ThCr$_2$Si$_2$ structure and has been investigated extensively due
to its pronounced non-Fermi liquid (NFL) properties at low $T$. This
behavior was attributed to a quantum critical point, which can be
attained by application of a small magnetic field of 60~mT within
the $ab$ plane.\cite{YRS-02-2} At intermediate temperatures
(10-300~K) the properties of YbRh$_2$Si$_2$ are determined by a
competition of crystal electric field (CEF) excitations and the
Kondo interaction. A knowledge of the corresponding characteristic
energy scales $k_\mathrm{B}T_{\mathrm{CEF}}$ and
$k_\mathrm{B}T_{\mathrm{K}}$ is essential for an understanding of
the low-temperature properties of the compound. The NFL behavior
dominates the thermodynamic and transport properties up to 10 K.
Notably, the electronic specific heat divided by $T$ exhibits a
logarithmic increase upon cooling with a spin fluctuation
temperature $T_0=24$~K.\cite{YRS-00-1} The corresponding entropy
revealed a doublet ground state of the Yb$^{3+}$ ions with a Kondo
temperature $T_\mathrm{K}$ of approximately 17 K.\cite{Thesis-7} The
CEF level scheme of YbRh$_2$Si$_2$ was determined from inelastic
neutron scattering: The $4f^{13}$ multiplet of the Yb$^{3+}$ is
split into 4 doublets at energies corresponding to 0-200-290-500~K,
respectively.\cite{YRS-06-4}

The strong interplay between Kondo effect and CEF splitting
manifests itself in the transport properties of the system. The
temperature dependencies of both resistivity \cite{YRS-05-4}
$\rho(T)$ and thermopower \cite{YRS-06-5} $S(T)$ exhibit a single
large extremum around 100~K, which was attributed to scattering on
the full Yb$^{3+}$ multiplet. No signature corresponding to Kondo
scattering on the ground-state doublet at $T_\mathrm{K}$ has been
found in $S(T)$ and $\rho(T)$. Such behavior was observed and
theoretically predicted for compounds, where the energy scales
$k_\mathrm{B}T_{\mathrm{CEF}}$ and $k_\mathrm{B}T_{\mathrm{K}}$ are
of the same order of magnitude, i.e. for systems near the crossover
from the HF to the intermediate-valent (IV)
regime.\cite{TB-99-2,TB-05-3,B-85-2}

Upon applying pressure, the Kondo temperature of Yb-based HF systems
is typically shifted to lower $T$, while the CEF levels are not
affected significantly. For sufficiently small values of
$T_{\mathrm{K}}$ separate maxima in $\rho (T)$ due to Kondo effect
on the ground state and on excited CEF levels are expected to occur.
Such behavior was confirmed for YbRh$_2$Si$_2$ by means of
resistivity investigations under pressure:\cite{YRS-05-4} Above 4
GPa, the single peak in $\rho (T)$ splits into three separate
maxima. The two maxima at lower $T$ were attributed to Kondo
scattering on the ground-state doublet with the onset of coherence
and to Kondo scattering on thermally populated CEF levels. The
origin of a third maximum remains unclear.

According to theoretical models, the lowering of $T_{\mathrm{K}}$ is
also expected to induce systematic changes in $S(T)$
(Ref.~\onlinecite{TB-05-3} and references therein): For systems with
a thermopower as described for YbRh$_2$Si$_2$ the anticipated
behavior upon application of a small pressure or weak lowering of
$T_{\mathrm{K}}$ is the appearance of a low-$T$ shoulder. With
further decreasing $T_{\mathrm{K}}$ two separate minima develop in
$S(T)$, similar to the behavior of $\rho (T)$. The minimum at lower
$T$, which reflects Kondo scattering on the ground-state doublet, is
situated at $T_{\mathrm{min 1}} \approx
T_\mathrm{K}$.\cite{TB-87-1,TB-97-1,TB-05-3} The high-$T$ minimum is
caused by Kondo scattering on thermally populated CEF levels. For an
excited CEF level at $\Delta _{\mathrm{CEF}}$ above the ground state
it typically appears at $T_{\mathrm{min 2}} \approx (0.3 \ldots 0.6)
\Delta _{\mathrm{CEF}} /
k_{\mathrm{B}}$.\cite{TB-76-1,TB-86-1,TB-05-3} Such evolution has
been detected e.g.~for Yb(Ni$_{x}$Cu$_{1-x}$)$_2$Si$_2$ upon
substitution (chemical pressure).\cite{TB-99-3} The IV system
YbCu$_2$Si$_2$ exhibits a single minimum in $S(T)$. As the Kondo
temperature is reduced by Ni substitution on the Cu site, a shoulder
appears at low $T$ already for the lowest Ni content studied ($x =
0.125$). However, it remains unsolved, whether only a single minimum
occurs, when the system is pushed to the HF regime for Ni
concentrations $x<0.125$, or whether the appearance of the low-$T$
shoulder is directly related to the development of the HF state.
YbRh$_2$Si$_2$ seems to be an appropriate system to address this
problem, since it, being a HF metal, exhibits a single minimum in
$S(T)$.

In order to change the effective coupling between the $4f$ and the
conduction electrons, substitution on all crystallographic sites has
been realized in YbRh$_2$Si$_2$, generally with the aim of lowering
the antiferromagnetic ordering temperature. This, however, is
connected with an increase in $T_\mathrm{K}$ as observed in
La$_{1-x}$Yb$_x$Rh$_2$Si$_2$,\cite{YRS-05-7} I-type
YbIr$_2$Si$_2$,\cite{YRS-05-12} and
YbRh$_2$(Si$_{1-y}$Ge$_y$)$_2$.\cite{YRS-02-3} Recent investigations
on Lu$_{1-x}$Yb$_x$Rh$_2$Si$_2$ indicate a weak lowering of
$T_\mathrm{K}$ upon replacement of Yb by the nonmagnetic
Lu.\cite{Sam} Additionally, the substitution on the Yb site leaves,
in a first approximation, the chemical environment of the remaining
Yb ions unchanged, thus reducing the influence of disorder on the
magnetic moments. Lu$_{1-x}$Yb$_x$Rh$_2$Si$_2$ therefore appeared a
promising candidate for the observation of distinct anomalies in the
thermopower due to Kondo interaction on the ground state and on
excited CEF levels at ambient pressure, with the objective to
determine the characteristic energy scales $k_\mathrm{B}
T_\mathrm{K}$ and $k_\mathrm{B} T_{\mathrm{CEF}}$ of pure
YbRh$_2$Si$_2$.

In this paper we present thermopower and resistivity measurements of
a number of Lu$_{1-x}$Yb$_x$Rh$_2$Si$_2$ single crystals with Yb
concentration $0 \leq x \leq 1$. The energy scales of Kondo
interaction and CEF excitations as a function of the Yb
concentration are deduced based on the thermopower and supported by
the resistivity data. Finally, the evolution of $S(T)$ and $\rho(T)$
is discussed in comparison to other Ce and Yb-based materials.

%%%%%%%%%%%%%%%%%%%%%%%%%%%%%%%%%%%%%%%%%%%%%%%%%%%%%%%%%%%%%%%%%%%%%%%%%%%%%%%%%%%%%%%%%%%

\section{Experimental details}

Single crystals of Lu$_{1-x}$Yb$_x$Rh$_2$Si$_2$ ($0 \leq x < 1$)
were grown from In flux, as described elsewhere.\cite{Sam} The
average (nominal) Yb content $x_{\mathrm{nom}}$ of each batch was
determined from susceptibility measurements and confirmed by
microprobe analysis on selected crystals. However, our resistivity
measurements suggest a moderate variation of the Yb concentration
throughout a batch. The scaling analysis described below indicates
deviations from $x_{\mathrm{nom}}$ of up to approximately 5 \%.
Specifically, the samples with \mbox{$x_{\mathrm{nom}} = 0.15$} and
\mbox{$x_{\mathrm{nom}} = 0.49$} studied in $S(T)$ and $\rho(T)$
have an effective Yb concentration of \mbox{$x = 0.10$} and \mbox{$x
= 0.44$}, respectively. For all other samples the nominal
concentration has been confirmed, i.e.~$x= x_\mathrm{nom}$. In the
following, the effective values, $x$, are used.

The lattice constants of the stoichiometric systems LuRh$_2$Si$_2$
and YbRh$_2$Si$_2$ were determined from X-ray diffraction
measurements on powdered material. The ThCr$_2$Si$_2$ crystal
structure has been confirmed within the doping series. No additional
peaks have been resolved in the pattern. The fraction of foreign
phases can thus be excluded to be higher than 2~\%. The microprobe
analysis has indicated no free elemental nor binary phases in the
studied samples. Due to the extremely small change of the unit cell
volume $V_{\mathrm{uc}}$ of \mbox{$0.41 \pm 0.12$ \%} with respect
to YbRh$_2$Si$_2$ a linear dependence of $V_{\mathrm{uc}}(x)$ is
assumed for the crystals with partial substitution. The tiny
variation in the lattice constant is not expected to significantly
influence the relative position of the CEF levels.

Investigations of the thermopower $S$ and the electrical resistivity
$\rho$ were performed within the $ab$ plane of the crystals with a
typical size of $4 \times 1 \times 0.05$ mm$^3$. Both quantities
were measured in the temperature range from 3 K to 300~K in a
commercial device (PPMS from Quantum Design) using the same
contacts. Measurements of $\rho$ were extended down to 0.4 K using a
$^3$He insert. The resistivity was determined with a 4-point ac
technique. For the thermopower a relaxation-time method with a
low-frequency square-wave heat pulse utilizing two thermometers was
used. The determination of $S$ implies an average over the contact
area of both, the voltage and the temperature gradient. In our
setup, due to the small crystal size, the contact size was not
negligible compared to the sample dimensions. However, data sets
obtained from repeated measurements on the same specimen but with
different contacts can be scaled on top of each other. In
particular, the position of the minimum in $S(T)$ remained
unaffected. The absolute values of $S$ could be reproduced within
\mbox{$\pm 8$ \%}. The thermopower data for the pure Yb compound are
taken from Ref.~\onlinecite{YRS-06-5}.

%%%%%%%%%%%%%%%%%%%%%%%%%%%%%%%%%%%%%%%%%%%%%%%%%%%%%%%%%%%%%%%%%%%%%%%%%%%%%%%%%%%%%%%%%%%

\section{Results}

\begin{figure}
\begin{center}
\includegraphics[bb=5 10 205 310, width=0.45\textwidth]{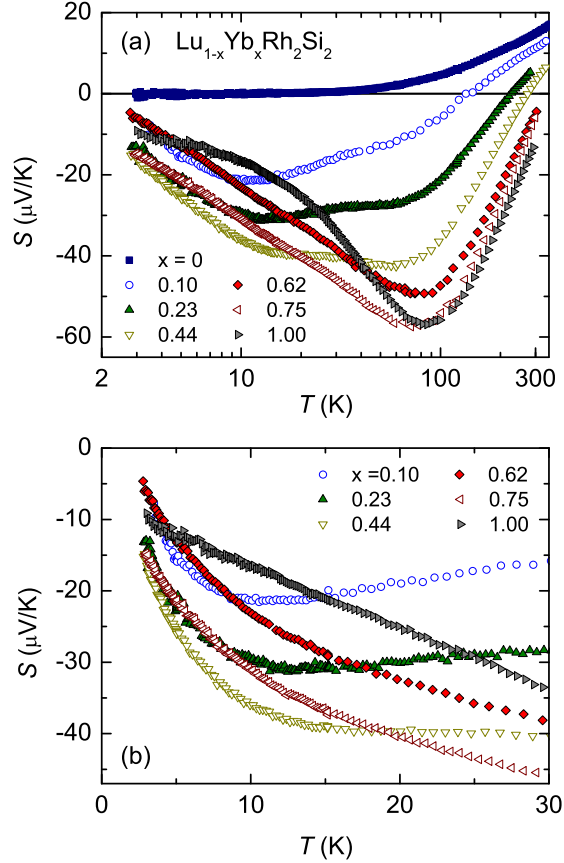}
\end{center}
\caption{(a) Temperature dependence of the thermopower of
Lu$_{1-x}$Yb$_x$Rh$_2$Si$_2$ ($0 \leq x \leq 1$). (b)
Low-temperature behavior of $S(T)$ of the crystals with $x > 0$.
\label{SvsT}}
\end{figure}

The thermopower $S(T)$ of Lu$_{1-x}$Yb$_x$Rh$_2$Si$_2$ \mbox{($0
\leq x \leq 1$)} is plotted semi-logarithmically in Fig.~\ref{SvsT}a
($x=0.08$ not shown for the sake of clarity). Fig.~\ref{SvsT}b
displays the low-$T$ behavior of the same curves on a linear
temperature scale. The thermopower of the reference compound
LuRh$_2$Si$_2$ is smaller than 1 $\mu$V/K in this $T$ range and
therefore omitted.

LuRh$_{2}$Si$_{2}$ exhibits a small positive thermopower, which is
typical for normal metals with hole-like charge carriers. The
thermopower comprises mostly a diffusion part of light nonmagnetic
charge carriers. A strong phonon drag contribution leading to an
enhancement around 20~K has not been resolved. By contrast, the
thermopower of YbRh$_{2}$Si$_{2}$ is negative in the whole
temperature range 3~K~$\leq T \leq$~300~K with large absolute
values. It shows a single broad minimum around 80~K, as typically
found in valence-fluctuating Yb compounds like
YbCu$_2$Si$_2$.\cite{TB-99-3} However, for the HF system
YbRh$_2$Si$_2$ the observed behavior was attributed to a combination
of Kondo interaction and CEF effects.\cite{YRS-06-5}
%Between 3 K and 40~K the thermopower of YbRh$_2$Si$_2$ shows an
%almost linear $S(T)$ dependence.
With decreasing Yb concentration the temperature dependence of the
thermopower changes qualitatively. The samples with $x=0.75$ and
$x=0.62$ exhibit a minimum at 80~K and a shoulder at low
temperatures, which may be seen on a linear scale
(Fig.~\ref{SvsT}b). For Yb concentrations of $x \leq 0.44$ the
thermopower minimum around 80~K clearly splits into 2 separate
features. While the position of the high-temperature shoulder
remains almost concentration independent, the low-temperature
shoulder shifts to lower $T$ upon further decreasing $x$.
Simultaneously the absolute values of the minimum structure at
elevated temperatures are significantly reduced. For samples with $x
\leq 0.44$ a sign change in $S(T)$ appears below room temperature,
which is shifted to lower $T$ with decreasing Yb concentration. This
indicates a stronger relative influence of the nonmagnetic
contribution to the thermopower in these samples.

\begin{figure}
\begin{center}
\includegraphics[width=0.45\textwidth]{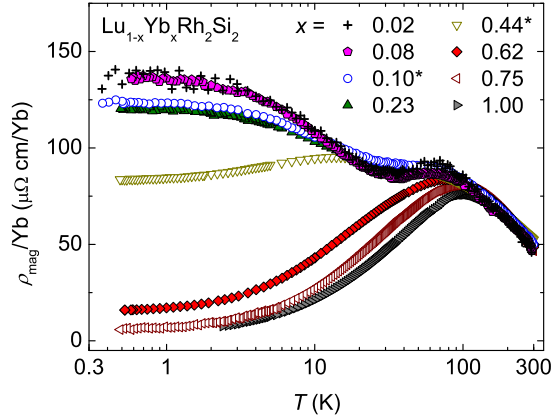}
\end{center}
\caption{Temperature dependence of the magnetic contribution to the
resistivity of \mbox{Lu$_{1-x}$Yb$_x$Rh$_2$Si$_2$} \mbox{($0 < x
\leq 1$)}. For samples denoted with an asterisk ($\star$) the
effective Yb concentration deviates from the nominal value, namely
$x = 0.44$ ($x_{\mathrm{nom}} = 0.49$) and $x = 0.10$
($x_{\mathrm{nom}} = 0.15$). \label{Rho}}
\end{figure}

The electrical resistivity of Lu$_{1-x}$Yb$_x$Rh$_2$Si$_2$ was
measured on the same samples as the thermopower. In addition, a
specimen with $x = 0.02$ has been investigated, which was too small
to measure $S(T)$. The results of $\rho (T)$ are in agreement with
Ref.~\onlinecite{Sam}. The resistivity of LuRh$_2$Si$_2$ takes a
value of about 30~$\mu \Omega$cm at 300~K and decreases linearly
from room temperature to 10~K, below which it reaches a constant
value of 1.25~$\mu \Omega$cm. The magnetic contribution,
$\rho_{\mathrm{mag}}$, was calculated by subtracting the data of the
reference compound LuRh$_2$Si$_2$ and a sample-dependent disorder
term $\rho^0_{\mathrm{dis}}$. The results normalized to the Yb
concentration are shown in Fig.~\ref{Rho}. For most samples scaling
of the data at elevated temperatures was achieved by using the value
of the nominal concentration. As already mentioned above, adjustment
of the effective concentration was necessary for $x_\mathrm{nom} =
0.15$ to $x=0.1$ and for $x_\mathrm{nom} = 0.49$ to $x = 0.44$ to
ensure scaling of the high-$T$ data above 100~K.

The magnetic resistivity $\rho _{\mathrm{mag}}(T)$ of the series
reflects the evolution from a diluted to a dense Kondo system, as
e.g.~demonstrated in Ce$_{x}$La$_{1-x}$Cu$_6$.\cite{B-87-2} At
temperatures $T > 100$~K, $\rho _{\mathrm{mag}}$ of all samples
increases logarithmically with decreasing $T$. Subsequently, samples
with low Yb concentrations $x \leq 0.23$ exhibit a plateau around 60
to 100~K, followed by a further increase in $\rho _{\mathrm{mag}}$
to lower $T$. At \mbox{$T < 4$ K} the magnetic resistivities of
these specimens saturate. On the other hand, the magnetic
resistivities of the concentrated Yb samples with $x \geq 0.62$ pass
through maxima around 70 to 100~K and then drop towards lower
temperatures. The plateau or maximum at elevated $T$ is attributed
to the presence of a CEF splitting in the system. The depopulation
of excited levels upon cooling and the associated lowering of the
scattering rate leads to a reduction of $\rho _{\mathrm{mag}}$ in
this temperature range. Towards lower $T$ the differences between
the diluted and the dense Yb systems become evident. While Kondo
scattering on the ground-state doublet gives rise to a second
increase as $-\ln T$ and a saturation at lowest $T$ for $x \leq
0.23$, the onset of coherence promotes a further decrease in the
magnetic resistivity in samples with $x \geq 0.62$. The sample with
Yb concentration $x = 0.44$ is situated close to the crossover
between the two regimes. It shows an only weak decrease in $\rho
_{\mathrm{mag}}$ towards low $T$ and a saturation at a relatively
large residual value.

%%%%%%%%%%%%%%%%%%%%%%%%%%%%%%%%%%%%%%%%%%%%%%%%%%%%%%%%%%%%%%%%%%%%%%%%%%%%%%%%%%%%%%%%%%%

\section{Discussion} %

For a quantitative analysis of the thermopower the magnetic
contribution, $S_{\mathrm{mag}}$, is usually determined by use of
the Gorter-Nordheim rule: $S_{\mathrm{mag}}\rho_{\mathrm{mag}} = S
\rho - S_{\mathrm{ref}}\rho_{\mathrm{ref}}$. $S_{\mathrm{ref}}$ and
$\rho_{\mathrm{ref}}$ are generally taken as the thermopower and
(total) resistivity of the nonmagnetic reference compound. The
resistivity $\rho_{\mathrm{ref}}$ may be approximated by a sum of a
phononic contribution $\rho^{\mathrm{ph}}_{\mathrm{ref}}$ and a
residual resistivity $\rho^{\mathrm{0}}_{\mathrm{ref}}$ due to
impurities. It is assumed that $\rho^{\mathrm{ph}}_{\mathrm{ref}}$
does not change significantly upon chemical substitution. The
disorder induced by doping, however, affects the residual
resistivity, and $\rho^0_{\mathrm{ref}}$ has to be replaced by the
$x$ dependent disorder contribution $\rho^0_{\mathrm{dis}}$ of the
alloys. Since $\rho^0_{\mathrm{dis}}(x)$ cannot be determined
accurately, an exact evaluation of $S_{\mathrm{mag}}$ is almost
impossible. For the presented data, the overall behavior of
$S_{\mathrm{mag}}$, and especially the position of the shoulders is
not expected to strongly deviate from that of $S$. At low
temperatures ($T<50$ K), at which the thermopower and the
resistivity of LuRh$_2$Si$_2$ are small, the difference between $S$
and $S_{\mathrm{mag}}$ is negligible. Just below room temperature,
the calculation of $S_\mathrm{mag}$ mainly implies a correction for
the diffusion thermopower of light charge carriers. The features
below 100K remain basically unchanged. For the discussion of the
data we therefore analyze $S \approx S_\mathrm{mag}$.

\begin{figure}
\begin{center}
\includegraphics[width=0.41\textwidth]{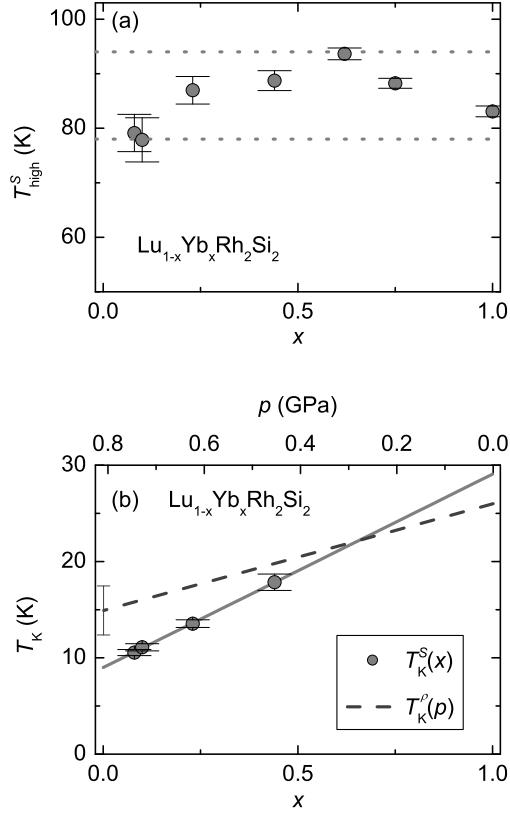}
\end{center}
\caption{Characteristic temperatures of Lu$_{1-x}$Yb$_x$Rh$_2$Si$_2$
at ambient pressure (a,b) and of YbRh$_2$Si$_2$ under pressure (b).
The x-axes in (b) are scaled to the same unit-cell volume. The
characteristic temperatures for $S(T)$ were obtained from a fit to
the data as explained in the text. The error bars represent the
uncertainty of the fitting procedure. The uncertainty in the
unit-cell volume, and consequently the pressure axis in (b) is
denoted by an error bar on $T_\mathrm{K}^{\rho}(p)$. The solid line
in (b) is a linear fit to $T_{\mathrm{low}}^S(x)$. \label{Minima}}
\end{figure}

Considering the Kondo and CEF energy scales of pure YbRh$_2$Si$_2$,
the two shoulders observed in the thermopower of
Lu$_{1-x}$Yb$_{x}$Rh$_2$Si$_2$ ($x<0.5$) are attributed to Kondo
scattering on the ground-state doublet and on thermally populated
multiplet states. Thus, the corresponding characteristic
temperatures $T_{\mathrm{low}}^S$ and $T_{\mathrm{high}}^S$ allow an
estimation of the evolution of $T_{\mathrm{K}}$ and
$\Delta_{\mathrm{CEF}}$ upon substitution. The position of the
low-$T$ minimum in the thermopower is usually close to the Kondo
temperature of the CEF ground state,\cite{TB-01-4} in agreement with
theoretical calculations for $s = 1/2$ neglecting CEF
effects.\cite{TB-87-1,TB-97-1} A recent calculation based on the
single-impurity Anderson model including both, a CEF splitting and
charge fluctuations yielded a similar result.\cite{TB-05-3} We
therefore assume \mbox{$T_{\mathrm{low}}^S = T_{\mathrm{K}}^S$}. The
high-temperature minimum is attributed to Kondo scattering from
thermally populated CEF levels. Thus, it represents a characteristic
temperature for the splitting of the relevant levels.
Fig.~\ref{Minima} shows the evolution of $T_{\mathrm{low}}^S$ and
$T_{\mathrm{high}}^S$ upon substitution. They have been determined
from a fit to the data between 4 and 200~K to a sum of two negative
Gaussian curves on a logarithmic temperature scale.

Fig.~\ref{Minima}a displays $T_{\mathrm{high}}^S$ vs.~effective Yb
content including the position of the single large minimum
$T_{\mathrm{min}}^S$ in the thermopower of pure YbRh$_2$Si$_2$. In
view of the substantial uncertainty in the determination of
$T_{\mathrm{high}}^S$, the position of the high-$T$ shoulder or
minimum in $S(T)$ may be regarded as temperature-independent. The
mean value of $86\pm 10$ K allows an estimate of the energetic
position of the relevant excited CEF levels. As theoretical
calculations predict $T_{\mathrm{high}}^S = (0.3 \ldots 0.6) \Delta
_{\mathrm{CEF}}/k_{\mathrm{B}}$,\cite{TB-76-1,TB-86-1,TB-05-3} the
observed minimum corresponds to a splitting of 150-290~K with
respect to the ground-state doublet. Thus, good agreement is
obtained with results of inelastic neutron scattering, which
revealed doublets at 0-200-290-500~K.\cite{YRS-06-4} We therefore
conclude that the large thermopower minimum around 80~K is caused by
Kondo scattering on the full Yb$^{3+}$ multiplet.

Fig.~\ref{Minima}b shows the position of the low-$T$ minimum
$T_{\mathrm{low}}^S = T_{\mathrm{K}}^S$ of the samples with $x<0.5$
vs.~$x$. For higher Yb concentrations a determination of
$T_{\mathrm{low}}^S$ was not possible with satisfactory precision.
The data sets for low $x$ clearly reveal an increasing
$T_{\mathrm{K}}^S$ with rising Yb concentration. A linear
extrapolation of $T_{\mathrm{low}}^S(x)$ yields a Kondo temperature
of 29 K for YbRh$_2$Si$_2$. The fit is shown as a line in
Fig.~\ref{Minima}b. This value is about a factor 1.5 larger than the
one obtained from the entropy of the system, namely
$T_{\mathrm{K}}^{c_P} = 17$ K.\cite{Thesis-7} However, the model
used in Ref.~\onlinecite{Thesis-7} for calculating $T_\mathrm{K}$
based on the entropy \cite{B-82-1} yields a temperature dependence
of $c_P$, which deviates significantly from that of the NFL compound
YbRh$_2$Si$_2$. Its application to this system is therefore somewhat
questionable. Furthermore, a determination of $T_\mathrm{K}$ from
different experimental probes usually yields different values,
however, always of the same order of magnitude.

The Lu-Yb substitution leads to a change in the unit cell volume
$V_{\mathrm{uc}}$. In order to evaluate the relevance of this effect
for the observed change in $T_{\mathrm{K}}$ a comparison to results
from experiments under pressure $p$ is of interest. Using the bulk
modulus \cite{YRS-03-3} of YbRh$_2$Si$_2$ of 189 GPa and assuming a
linear relation of $V_{\mathrm{uc}}$ vs.~$x$ we can compare our
results with investigations under pressure. The Kondo temperature
$T_{\mathrm{K}}^{\rho}(p)$ of YbRh$_2$Si$_2$ determined from
resistive pressure studies \cite{Thesis-5} is shown as a dashed line
in Fig.~\ref{Minima}b, where the pressure axis is scaled to the same
unit-cell volume change. It is seen that the lowering of $T_K$ under
pressure is somewhat weaker than upon substitution. This reflects
the relevance of additional effects beside the variation of
$V_{\mathrm{uc}}$. Chemical substitution induces a change in the
band structure of the system, which may influence the effective
coupling between the $4f$ and conduction electrons even at constant
$V_{\mathrm{uc}}$. However, the relatively small difference between
the evolution of $T_\mathrm{K}$ under pressure and upon Lu
substitution underlines a dominating influence of the unit cell
volume in Lu$_{1-x}$Yb$_x$Rh$_{2}$Si$_{2}$. This is ascribed to the
unchanged chemical environment of the $4f$ moments as a result of
the substitution on the rare-earth ion site.
%This may be caused by the substitution on the rare-earth ion site
%since it leaves the chemical environment of the $4f$ moments in
%first approximation unchanged.
Furthermore, due to the small radius of the $4f$ shell, Lu and Yb
behave chemically very similar. The $4f$ electrons together with the
nucleus act as an "effective nucleus". Therefore, Lu-Yb substitution
is expected to have a minor influence on the band structure. Yet,
the integer valence $v$ of the Lu ions compared to the slightly
reduced value for the Yb ions (Lu$^{3+}$ vs.~ Yb$^{v+}$ with $2.95
\leq v\leq 3.00$)\cite{Mitt-6} might have a small effect on the
charge-carrier concentration.

The large values usually found for the thermopower of $4f$ systems
have been related to the likewise enhanced electronic contributions
to the specific heat. In the zero temperature limit, the ratio
$S/\gamma T$ of several correlated compounds takes a quasi-universal
value.\cite{TB-04-2} For metals the dimensionless quantity $q
=N_{\mathrm{AV}}e S / \gamma T $ with the Avogadro number
$N_{\mathrm{AV}}$ and the electron charge $e$ is close to $\pm 1$,
whereas the sign depends on the type of charge carriers. This
relation can be derived within Fermi-liquid theory assuming impurity
scattering as the relevant scattering process.\cite{TB-05-4} In the
present system a thermopower $S \propto T$ is only found for $T < 5$
K and $x < 0.5$, due to both the low Kondo temperature of the order
of 10 K and the NFL behavior of the pure system YbRh$_2$Si$_2$. The
calculated $q$ values ranging from $-0.63$ to $-0.85$ are in line
with that expected for hole-like charge carriers of~$-1$.

The change in the behavior of $S(T)$ from a single minimum at $x=1$
to a double-peak structure for small $x$ is correlated to the
lowering of $T_\mathrm{K}$ upon decreasing Yb concentration.
YbRh$_{2}$Si$_{2}$ with a Kondo temperature of approximately 20~K
and a first excited CEF level around 200~K seems to be situated near
the critical ratio $T_\mathrm{CEF}/T_\mathrm{K}$ where the two
minima of Kondo scattering on ground state and thermally populated
CEF levels in $S(T)$ merge into a single feature. A slight reduction
of $T_\mathrm{K}$ on the other hand allows for a separation of both
effects.

\begin{figure}
\begin{center}
\includegraphics[bb=5 10 230 285, width=0.45\textwidth]{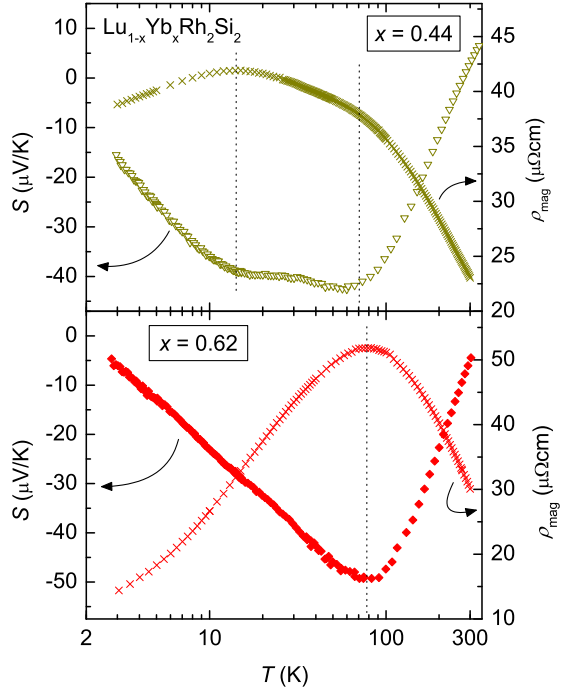}
\end{center}
\caption{Comparison between thermopower and magnetic contribution to
the resistivity for two Yb concentrations $x = 0.44$ and $x = 0.62$.
\label{SVglRho}}
\end{figure}

A similar behavior is found in the magnetic contribution to the
resistivity $\rho_{\mathrm{mag}}(T)$ of
Lu$_{1-x}$Yb$_x$Rh$_2$Si$_2$. The similarities are best seen for the
two samples with Yb concentrations at the crossover from a single
large minimum in $S(T)$ to two clearly separated shoulders.
Fig.~\ref{SVglRho} shows a comparative plot of $S(T)$ as well as
$\rho_{\mathrm{mag}}(T)$, which were determined on the same
specimens with $x = 0.44$ and 0.62. The curves for thermopower and
magnetic resistivity strongly resemble each other. For the sample
with $x = 0.44$ two clearly separated shoulders are seen in both
quantities, situated at the same temperatures as indicated by the
vertical lines. By contrast, the sample with $x=0.62$ exhibits only
one large extremum in $S(T)$ and $\rho_{\mathrm{mag}}(T)$ around
80~K. The double-peak structure is well tracked by both the
thermopower and resistivity for $x < 0.5$, whereas larger
concentrations $x > 0.5$ exhibit a single peak.

It appears remarkable that the effects of Kondo scattering on the
ground-state doublet and on thermally populated CEF levels cannot be
separated for $x>0.5$ in $S(T)$ and $\rho(T)$, although the
corresponding energy scales differ by approximately one order of
magnitude. This observation can be understood in view of the
relatively large temperature range, which is effected by the thermal
population of an excited CEF level as apparent e.g.~from a Schottky
contribution to the specific heat. A significant population of an
excited CEF level and, consequently, an enhanced scattering from CEF
excitations sets in at temperatures well below the splitting of
$\Delta_{\mathrm{CEF}}/k_{\mathrm{B}}$. Thus, in
Lu$_{1-x}$Yb$_x$Rh$_2$Si$_2$ for $x>0.5$ the crossover from
scattering solely on the ground-state doublet to that on the all CEF
level may not be resolved. Instead, the observed single large
extremum in $S(T)$ and $\rho(T)$ is caused by scattering on the full
Yb$^{3+}$ multiplet. On the other hand, in samples with $x<0.5$ the
reduced Kondo scale allows a separate observation of scattering on
the ground-state doublet at low $T$, while thermal population of the
three excited levels give rise to a second broad extremum at
elevated temperatures.

The disappearance of the low-$T$ minimum in the thermopower of
Yb(Ni$_{x}$Cu$_{1-x}$)$_2$Si$_2$ for $x = 0$ is connected with a
crossover to the valence-fluctuating regime.\cite{TB-99-3} A
qualitatively similar behavior has been frequently observed in Ce
systems under pressure and upon substitution, e.g.~in CeRu$_2$Ge$_2$
\cite{TB-04-1} or Ce(Ni$_x$Pd$_{1-x}$)$_2$Si$_2$.\cite{TB-01-3} In
contrast to Yb compounds, Ce based HF systems generally exhibit
maxima in $S(T)$ due to the preponderance of electron-like charge
carriers. Likewise, the crossover from two maxima to a single
maximum is usually taken as an indication \cite{TB-04-1,TB-01-3}
that $T_\mathrm{K}$ lies in the order of the CEF splitting and the
system enters the valence-fluctuating regime. However, in XANES
measurements of the L-III absorption edge of the Yb ion in
Lu$_{1-x}$Yb$_x$Rh$_2$Si$_2$ no significant contribution from
Yb$^{2+}$ could be resolved at 5~K. Taking into account the
resolution of the method, the valence $v$ has been estimated to be
$2.95 \leq v \leq 3.00$.\cite{Mitt-6} A strong intermediate-valent
character is therefore excluded for this system. Thus,
Lu$_{1-x}$Yb$_x$Rh$_2$Si$_2$ appears to be a rare example, for which
the two extrema in $S(T)$ merge, while the system is still in the HF
regime. The coalescence of both features corresponding to
$T_{\mathrm{K}}$ and $\Delta_{\mathrm{CEF}}$ can be observed in
detail in this series since the Lu substitution induces an extremely
small change of the unit-cell volume connected with a very weak
lowering of $T_\mathrm{K}$.

%%%%%%%%%%%%%%%%%%%%%%%%%%%%%%%%%%%%%%%%%%%%%%%%%%%%%%%%%%%%%%%%%%%%%%%%%%%%%%%%%%%%%%%%%%%

\section{Summary}
The temperature dependence of the thermopower of
Lu$_{1-x}$Yb$_x$Rh$_2$Si$_2$ qualitatively changes upon substitution
from a single large minimum in $S(T)$ for the pure Yb compound to
two well separated minima for $x < 0.5$. A similar evolution is
found in the magnetic contribution to the resistivity. This change
in the overall behavior of $S(T)$ and $\rho _{\mathrm{mag}}(T)$ is
ascribed to a lowering of $T_\mathrm{K}$ upon decreasing Yb content.
For high Yb concentrations $x > 0.5$ the extrema of Kondo scattering
on the ground state and thermally populated CEF levels merge into
one single feature. A slight reduction of $T_\mathrm{K}$ due to the
substitution of Yb by Lu on the other hand allows for a separation
of both effects in the transport properties of the series for $x <
0.5$. The evolution of the Kondo temperature upon substitution can
be understood mainly from the change in the unit-cell volume. In
addition, modifications in the band structure may be relevant. Due
to the extremely small overall change in $T_\mathrm{K}$,
Lu$_{1-x}$Yb$_x$Rh$_2$Si$_2$ displays the crossover from two minima
in $S(T)$ to one minimum upon increasing $x$ without entering the
valence-fluctuating regime. The Kondo temperature of YbRh$_2$Si$_2$
has been estimated to be around 29~K.

\begin{acknowledgments}
We are grateful to V.~Zlati\'{c}, S.~Burdin, and C.~Geibel for
stimulating discussions. We thank N. Caroca-Canales for X-ray
diffraction measurements on LuRh$_2$Si$_2$. UK acknowledges
financial support by COST action P16. Work at UC Irvine (SM and ZF)
has been supported by NSF Grant No. DMR-0710492.
\end{acknowledgments}

\end{document}